\documentclass[12pt]{article}
\usepackage[dvips]{color}
\usepackage{epsfig}
\usepackage{graphicx}

\begin{document}
\setcounter{page}{1}
\pagestyle{plain} \vspace{1cm}
\begin{center}
\Large{\bf Deformation of the quintom cosmological model and its consequences}\\
\small \vspace{1cm} {\bf J. Sadeghi \footnote{pouriya@ipm.ir}},\quad{\bf B. Pourhassan
\footnote{b.pourhassan@du.ac.ir}},\quad {\bf Z. Nekouee
\footnote{z.nekouee@stu.umz.ac.ir}} and\quad{\bf M. Shokri \footnote{mehdi.shokri@uniroma1.it}}\\
\vspace{0.5cm}$^{1}${\it Department of Physics, Faculty of Basic Sciences,\\
University of Mazandaran, P. O. Box 47416-95447, Babolsar, Iran}\\
\vspace{0.5cm}$^{2}${\it School of Physics, Damghan University, Damghan, 3671641167, Iran}\\
\vspace{0.5cm}$^{3}${\it Department of Mathematic, Faculty of Mathematic,\\
University of Mazandaran, P. O. Box 47416-95447, Babolsar, Iran}\\
\vspace{0.5cm}$^{4}${\it Physics Department and INFN,\\
Universit`a di Roma "La Sapienza", Ple. Aldo Moro 2, 00185, Rome, Italy}\\
\end{center}\vspace{1.2cm}
\begin{abstract}
In this paper, we  investigate the effects of non-commutative phase-space on the quintom cosmological model. In that case, we discuss about some cosmological parameters and show that they are depend on the deformation parameters. We find that the non-commutative parameter plays important role which help to re-arrange the divergency of cosmological constant. We draw time-dependent scale factor and investigate the effect of non-commutative parameters. Finally, we take advantage from non-commutative phase-space and obtain the deformed Lagrangian for the quintom model. In order to discuss some cosmological phenomena as dark energy and inflation, we employ Noether symmetry.
\end{abstract}
\newpage
\section{Introduction}
More than  one decade, the dark energy (DE) is discovered as the
late time acceleration of the universe, which confirmed by several
important observations \cite{1,2,3,4}. However, the nature of this phenomena
is unknown, but there are some proposals to describe it. For instance modification of general relativity can help us to
clarify the DE subject. This modification can be done by some
changes in the Einstein field equation. If we focus on the right
hand side of the Einstein equation, limit ourselves to the modified
theories of gravity as $\emph{f}(R)$
theories \cite{S1, S2, S3, f1, f2, f3, f4}, extra dimensions, and scalar-tensor theories. On the other hand, it is believed that we can dissolve
the DE mystery by adding some matters to the components of the
universe instead of modification gravity. In this approach the equation state as $w=-1$ is one of the most significant candidates to
illustrate DE \cite{5,6,7,8}. However, the cosmological constant suffers from some problems, one of them is that there is
a big discrepancy between the value of cosmological constant that
is proposed by fundamental physics and its observational value
\cite{9}. Another alternative for DE is quintessence with $-1\leq
w\leq1$ \cite{10, 10-1,10-2, 10-3, 11, P2, P3}, which is a real scalar field that changes from one
space-time point to another, unlike the cosmological constant that
is uniformly constant. Also, it can be either attractive or
repulsive depending on the ratio of its kinetic and potential
energy. According to the observational evidences, we know that in order to
have accelerating expansion, the equation of state parameter have to be
$w<-\frac{1}{3}$. Because of this reality, people are motivated to
consider another type of matter that is called phantom field with
$w<-1$ \cite{S11, S22, S33, S44, S55, 12, 12-3, 12-4}. The remarkable feature for phantom is that it
possesses a negative kinetic energy. In that case, it could cause
the expansion of the universe to accelerate so quickly that as known as the Big Rip scenario. Another interesting model of dark energy which is an alternative to quintessence is based on the Chaplygin gas equation of state \cite{34}. Origin of this model in cosmology inspired by string theory \cite{36, 37}. For some theoretical and observational point of view, generalized Chaplygin gas \cite{S111, 35, 38} and viscous generalized Chaplygin gas \cite{39,40,41,42,43} equation of states emerged. For the same reasons, modified Chaplygin gas \cite{44,44-1} and viscous modified Chaplygin gas \cite{45, 46} equation of state proposed to describe dark energy and also dark matter. A further extension of Chaplygin gas
model is called modified cosmic Chaplygin gas \cite{47, 48} or newest extended Chaplygin gas \cite{66,67,68,69,70,71}. which could recover barotropic fluid.\\

In addition to the mentioned DE models, another scenario for DE is
quintom with the equation of state crossing the cosmological
constant boundary $w=-1$ \cite{13}. One can see Ref. \cite{new1} for the comprehensive review on quintom cosmology. Theoretically, at least two
degrees of freedom are required to make this DE model. Hence,
typically we consider two scalar fields for quintom cosmological
models, one of them is a normal scalar field and the other is a
phantom-type scalar field. In that case, Ref. \cite{new2} studied phase space of two-field quintom models of dark energy and found that quintom cosmology is self-dual. We note that among all above DE models,
the present work is based on quintom models which is one of the interesting scenarios in cosmology. There are important application in the early universe, for example, nonsingular bouncing solution obtained from quintom cosmology \cite{new3, new4}. Also, it has relation with string theory \cite{new5} which make it strong model in cosmology.

As we know there are many physical theories that are based to
non-commutative phase-space approach \cite{N1, N2, N3}.  The cosmological scenario will be interested under deformation of mini-superspace. In that case, there are many attempts to use non-commutative phase-space approach in cosmology \cite{14, 15, 16, 17, 18}. For  Kaluza-Klein cosmology in (4+1)-dimension with
a negative cosmological constant, people were interested to solve the Hierarchy problem by non-commutative mini-superspace \cite{15}.
Also, the deformed phase - space by non-commutativity has been
analyzed for Kaluza-Klein cosmology \cite{18}. There are some works to
investigate non-commutative mini-superspace on scalar field
cosmology in minimal and non-minimal cases like \cite{14, 17}.
So, here  we are concerning  to apply this kind of
non-commutative for quintom system to illuminate
hidden aspects of the model. specially our model may be examined by the CMB observations because the noncommutativity could lead to testable signals in the power spectrum of primordial gravitational waves \cite{new6} which yields to testing quantum gravity effects with latest CMB observations \cite{new7}.

Here, in order to clarify the
non-commutative effects on the corresponding system, we focus on the cosmological constant. Effect of non-commutative space on the cosmological scenarios are interesting subject of study. For example, the effects of non-commutativity in the phase-space of the classical and quantum cosmology of
Bianchi models are investigated by the Ref. \cite{vakili}. All above information give us motivation to
arrange paper as follow. In section 2, we introduce some DE models
that are based on single scalar field with different equation of
state parameter,  and also we review their configurations in
context of non-commutative mini-superspace. In section 3, the
quintom model is introduced and some new variables are defined
that are related to our main purpose. In section 4 we present the non-commutative formalism of the model by
deformation of mini-superspace. In that section,
also we have some figures about behavior of scale factor with
respect to time and discuss three special cases for
non-commutative parameters. In section 5, we use non-commutative
phase space and write the corresponding Lagrangian for the quintom
model. Moreover, in order to discuss the inflation and dark
energy, we use Noether symmetry approach. Eventually, conclusion
and discussion about the
obtained results of this scenario is expressed in section 6.
\section{The Deformed Phase Space of Single Scalar Field}
As we told in the previous section, in order to explain DE, we can adopt
our model on single scalar field scenarios with different equation
of state parameters such as the cosmological constant,
quintessence and phantom energy. On the other hand, scalar field in
both cases of minimal and non-minimal play a great role in
cosmological theories like inflation and dark matter. The
deformation of phase space on the scalar field can be clarified by
other aspects of corresponding theories. Because of this fact, in
this section, we would like to review some important results about deformation of phase space in cosmological point of
view.\\

In the Ref. \cite{19} the deformed phase space of
(4+1)-dimensional Kaluza-Klein cosmology with a negative
cosmological constant has been studied. They have shown that the effective cosmological
constant depends on the deformation parameters $\theta$ and
$\beta$. In the Ref. \cite{18} the authors used the non-commutative
mini-superspace of Kaluza-Klein cosmology and investigated the
late time acceleration. Also, the issue has been studied to
dilaton cosmology for both cases commutative and non-commutative \cite{16}. The results presented that in commutative case, the scale
factor has a decelerated expansion in the early universe. For the late
time, the universe experiences the power-law expansion which is not
consistent with observations. But in non-commutative case, the scale factor undergo an exponential expansion in late time which is acceptable from observational point of view. Ref. \cite{17} have been shown that the model of deformed phase space on the single scalar field  behaves like a de Sitter universe at the late time, also the effective cosmological constant related to
the non-commutative parameter. The application of non-commutative
mini-superspace for the non-minimal case of scalar field is also
considered in the Ref. \cite{14}, and it is compared with its commutative
counterpart. So, generally we can say that the parameters of
non-commutative affect on the stability and also cosmological
constant. Also it is possible to study the behavior of gravitational wave background corresponding to the inflationary age in the non-commutative field approach to test non-commutative aspect of nature \cite{new6}. In that case it is found that in quantum gravitational theory, the non-commutative phase space
leads to a modified dispersion relation of gravitational waves \cite{new7} which is important from cosmological point of view. For these reasons, we are going to consider
special system and see the effects of non-commutative parameters on the corresponding system.
\section{The Quintom  Model in Commutative Phase Space }
We focus on a typically quintom model with two scalar fields, the first is a canonical scalar field $\phi$ and the second is a
phantom field $\sigma$ with negative kinetic energy. It is indeed another kind of two-component fluid models which considered in literatures like \cite{Kh111}. The action of this case takes the following form,
\begin{equation}\label{1}
S=\frac{1}{2}\int
d^{4}x\sqrt{-g}\left[R-g^{\mu\nu}\partial_{\mu}\phi\partial_{\nu}\phi+g^{\mu\nu}\partial_{\mu}\sigma\partial_{\nu}\sigma+V(\phi,\sigma)\right],
\end{equation}
where $V(\phi,\sigma)$ is the potential function of model and also
we use the cosmological unit ($8\pi G=c=1$). In order to
analysis the model, we consider the FRW metric with the line element as,
\begin{equation}\label{2}
ds^{2}=-dt^{2}+a^{2}(t)[dr^{2}+r^{2}d\Omega^{2}],
\end{equation}
where $d\Omega^{2}=d\theta^{2}+\sin^{2}\theta d\varphi^{2}$. By
using the metric (\ref{2}) in the action (\ref{1}) one can obtain the effective lagrangian as follow,
\begin{equation}\label{3}
\ell=-3\dot{a}^{2}a+a^{3}(\frac{1}{2}\dot{\phi}^{2}-\frac{1}{2}\dot{\sigma}^{2}-V(\phi,\sigma)).
\end{equation}

As we know the terms of potential play an important role in the action. Here, we consider two corresponding fields as $\phi$ and $\sigma$. Generally, the form of potential must be as follow,
\begin{equation}\label{4}
V(\phi,\sigma)=V(\phi)+V(\sigma)+V(\phi)V(\sigma).
\end{equation}
In order to modify the system, the Eq. (\ref{4}) for the potential will be complicated. Because of this reason, if we want to consider the Eq. (\ref{4}), we have to apply the formula $\{f,g\}_{\alpha}=f\star_{\alpha}g-g\star_{\alpha}f$ for the third term of potential. This leads us to some complicated calculations and also we have to care about some cosmological constants. For simplicity and some physical reasons, we restrict ourselves to the non-interacting case with the
following form of potential,
\begin{equation}\label{5}
V(\phi,\sigma)=V(\phi)+V(\sigma).
\end{equation}

In order to have the corresponding Hamiltonian from the Eq. (\ref{3}) in the form of harmonic oscillator, one can consider following changes of variables
\begin{equation}\label{6}
x_{1}=\frac{4}{3}a^{\frac{3}{2}}\sinh(\frac{3}{4\sqrt{2}}\phi),\quad
y_{1}=\frac{4}{3}a^{\frac{3}{2}}\cosh(\frac{3}{4\sqrt{2}}\phi),
\end{equation}
and
\begin{equation}\label{7}
x_{2}=\frac{2}{3}a^{\frac{3}{2}}\cosh(\frac{3}{2\sqrt{2}}\sigma),\quad
y_{2}=\frac{2}{3}a^{\frac{3}{2}}\sinh(\frac{3}{2\sqrt{2}}\sigma).
\end{equation}
Hence, the Hamiltonian calculated as follow,
\begin{equation}\label{8}
H=[({p_{x_{1}}}^{2}-{p_{y_{1}}}^{2})+{\omega_{1}}^{2}({x_{1}}^{2}-{y_{1}}^{2})]+[({p_{x_{2}}}^{2}-{p_{y_{2}}}^{2})+{\omega_{2}}^{2}
({x_{2}}^{2}-{y_{2}}^{2})],
\end{equation}
where
\begin{equation}\label{9}
p_{x_{1}}=2\dot{a}a^{\frac{1}{2}}\sinh(\frac{3}{4\sqrt{2}}\phi)+\frac{\sqrt{2}}{2}\dot{\phi}a^{\frac{3}{2}}\cosh(\frac{3}{4\sqrt{2}}\phi),\\
\end{equation}
\begin{equation}\label{10}
p_{y_{1}}=2\dot{a}a^{\frac{1}{2}}\cosh(\frac{3}{4\sqrt{2}}\phi)+\frac{\sqrt{2}}{2}\dot{\phi}a^{\frac{3}{2}}\sinh(\frac{3}{4\sqrt{2}}\phi),\\
\end{equation}
\begin{equation}\label{11}
p_{x_{2}}=\dot{a}a^{\frac{1}{2}}\cosh(\frac{3}{2\sqrt{2}}\sigma)+\frac{\sqrt{2}}{2}\dot{\sigma}a^{\frac{3}{2}}\sinh(\frac{3}{2\sqrt{2}}\sigma),\\
\end{equation}
\begin{equation}\label{12}
p_{y_{2}}=\dot{a}a^{\frac{1}{2}}\sinh(\frac{3}{2\sqrt{2}}\sigma)+\frac{\sqrt{2}}{2}\dot{\sigma}a^{\frac{3}{2}}\cosh(\frac{3}{2\sqrt{2}}\sigma),\\
\end{equation}
and
\begin{equation}\label{13}
{\omega_{1}}^{2}=-\frac{9\Lambda_{1}}{16},\quad
{\omega_{2}}^{2}=\frac{9\Lambda_{2}}{4}.
\end{equation}

It is obvious that the Hamiltonian given by the Eq. (\ref{8}) has an oscillator form that
is useful for gravitational theories. If we are interested to
discuss about commutative case of model, we can work with the usual
Poisson brackets which are given by,
\begin{equation}\label{14}
\{x_{i},x_{j}\}=0,\quad\{P_{x_{i}},P_{x_{j}}\}=0,\quad\{x_{i},P_{x_{j}}\}=\delta_{ij},
\end{equation}
where $x_{i}(i=1,2)$ and $p_{i}(i=1,2)$. We notice that by using
the Wheeler-Dewitt equation, we can access to the quantum
cosmology of the model. In order to compare non-commutative and
commutative phase spaces, firstly we calculate the quantities in
commutative space. Therefore, we try to obtain $x_{1}$,
$x_{2}$, $y_{1}$, $y_{2}$ and their derivatives with respect to
$\omega_{1}$ and $\omega_{2}$ which are given by,
\begin{equation}
\begin{array}{cc}
  \dot{x_{1}}=2p_{x_{1}}, & \dot{y_{1}}=-2p_{y_{1}},\\
  \dot{p_{x_{1}}}=-2{\omega_{1}}^{2}x_{1}, & \dot{p_{y_{1}}}=2{\omega_{1}}^{2}y_{1},\\
\end{array}
\end{equation}
and
\begin{equation}
\begin{array}{cc}
  \dot{x_{2}}=2p_{x_{2}}, & \dot{y_{2}}=-2p_{y_{2}},\\
  \dot{p_{x_{2}}}=-2{\omega_{2}}^{2}x_{2}, & \dot{p_{y_{2}}}=2{\omega_{2}}^{2}y_{2},\\
\end{array}
\end{equation}
so the corresponding  $x_{1}$, $x_{2}$,  $y_{1}$ and $y_{2}$ will be,
\begin{equation}
\begin{array}{cc}
  x_{1}(t)=\mu_{1}\sin(2\omega_{1}t+\varphi_{1}), & y_{1}(t)=\nu_{1}\sin(2\omega_{1}t+\psi_{1}),\\
  x_{2}(t)=\mu_{2}\sin(2\omega_{2}t+\varphi_{2}), & y_{2}(t)=\nu_{2}\sin(2\omega_{2}t+\psi_{2}).\\
\end{array}
\end{equation}

Finally, we are going to calculate the scale factor and fields of $\phi$ and $\sigma$. In order to obtain such quantities we use Eqs. (6-13) and (15-17), so one can obtain scale factor and fields as,
\begin{equation}
\begin{array}{cc}
  a^{3}(t)=\frac{9}{16}[\nu_{1}^{2}\sin^{2}(2\omega_{1}t+\psi_{1})-\mu_{1}^{2}\sin^{2}(2\omega_{1}t+\varphi_{1})],\\
  a^{3}(t)=\frac{9}{4}[\mu_{2}^{2}\sin^{2}(2\omega_{2}t+\varphi_{2})-\nu_{2}^{2}\sin^{2}(2\omega_{2}t+\psi_{2})],\\
\end{array}
\end{equation}
and
\begin{equation}
\begin{array}{cc}
\phi(t)=\frac{2\sqrt{2}}{3}\ln\frac{(\nu_{1}\sin(2\omega_{1}t+\psi_{1})+\mu_{1}\sin(2\omega_{1}t+\varphi_{1}))^{2}}{\nu_{1}^{2}\sin^{2}(2\omega_{1}t+\psi_{1})-\mu_{1}^{2}\sin^{2}(2\omega_{1}t+\varphi_{1})},\\
\sigma(t)=\frac{\sqrt{2}}{3}\ln\frac{(\nu_{2}\sin(2\omega_{2}t+\psi_{2})+\mu_{2}\sin(2\omega_{2}t+\varphi_{2}))^{2}}{\mu_{2}^{2}\sin^{2}(2\omega_{2}t+\varphi_{2})-\nu_{2}^{2}\sin^{2}(2\omega_{2}t+\psi_{2})},\\
\end{array}
\end{equation}
where $\mu_{1}$, $\mu_{2}$, $\nu_{1}$, $\nu_{2}$, $\psi_{1}$, $\psi_{2}$, $\varphi_{1}$ and $\varphi_{2}$ are arbitrary constants. Later, we will discuss about behavior of the scale factor and show that is increasing function of time.
\section{The Quintom Model in Non-commutative Phase Space}
In canonical quantum cosmology, the Wheeler-DeWitt equation is used
for quantization of mini-superspace variables. There is another
viewpoint for quantization of the mini-superspace that is
established on the deformation of phase space of the model. In this
approach, quantum effects can be dissolved by the Moyal brackets
$\{f,g\}_{\alpha}=f\star_{\alpha}g-g\star_{\alpha}f$ which is based
on the Moyal product as,
\begin{equation}\label{20}
(f\star_{\alpha}g)(x)=exp[\frac{1}{2}\alpha^{ab}\partial_{a}^{(1)}\partial_{b}^{(2)}]f(x_{1})g(x_{2})|_{x_{1}=x_{2}=x}.
\end{equation}
After corresponding calculations, we find the algebra of variables
as,
\begin{equation}\label{21}
\{x_{i},x_{j}\}_{\alpha}=\theta_{ij},\quad\{x_{i},P_{j}\}_{\alpha}=\delta_{ij}+\sigma_{ij},\quad\{P_{i},P_{j}\}=\beta_{ij}.
\end{equation}
Transformations on the classical phase space variables are expressed
as,
\begin{equation}\label{22}
\hat{x}_{1}=x_{1}+\frac{\theta}{2}p_{y_{1}},\quad \hat{y}_{1}=y_{1}-\frac{\theta}{2}p_{x_{1}},\quad \hat{p}_{x_{1}}=p_{x_{1}}-\frac{\beta}{2}y_{1},
\quad \hat{p}_{y_{1}}=p_{y_{1}}+\frac{\beta}{2}x_{1},\\ \\
\end{equation}
and
\begin{equation}\label{23}
\hat{x}_{2}=x_{2}+\frac{\theta}{2}p_{y_{2}},\quad \hat{y}_{2}=y_{2}-\frac{\theta}{2}p_{x_{2}},\quad
\hat{p}_{x_{2}}=p_{x_{2}}-\frac{\beta}{2}y_{2},\quad \hat{p}_{y_{2}}=p_{y_{2}}+\frac{\beta}{2}x_{2}.\\
\end{equation}
The deformed algebra for new variables are given by,
\begin{equation}\label{24}
\{\hat{y},\hat{x}\}=\theta,\quad\{\hat{x},\hat{P_{x}}\}=\{\hat{y},\hat{P_{y}}\}=1+\sigma,\quad\{\hat{P_{y}},\hat{P_{x}}\}=\beta,
\end{equation}
where $\sigma=\frac{\beta\theta}{2}$. In order to construct the
deformed formalism of the model, we borrow the form of Hamiltonian
from Eq. (\ref{11}) with new variables in Eq. (18). Hence, the form of
Hamiltonian in the deformed analysis is found as,
\begin{eqnarray}\label{25}
H&=&[({p_{x_{1}}}^{2}-{p_{y_{1}}}^{2})+{\tilde{\omega_{1}}}^{2}({x_{1}}^{2}-{y_{1}}^{2})-{\gamma_{1}}^{2}
(x_{1}p_{y_{1}}+p_{x_{1}}y_{1})]\nonumber\\
&+&[({p_{x_{2}}}^{2}-{p_{y_{2}}}^{2})+{\tilde{\omega_{2}}}^{2}({x_{2}}^{2}-{y_{2}}^{2})-{\gamma_{2}}^{2}(x_{2}p_{y_{2}}+p_{x_{2}}y_{2})],
\end{eqnarray}
where
\begin{equation}\label{26}
{\tilde{\omega_{1}}}^{2}=\frac{{\omega_{1}}^{2}-\frac{\beta^{2}}{4}}{1-{\omega_{1}}^{2}
\frac{\theta^{2}}{4}},\quad
{\gamma_{1}}^{2}=\frac{\beta-{\omega_{1}}^{2}\theta}{1-{\omega_{1}}^{2}\frac{\theta^{2}}{4}},\quad
{\tilde{\omega_{2}}}^{2}=\frac{{\omega_{2}}^{2}-\frac{\beta^{2}}{4}}{1-{\omega_{2}}^{2}\frac{\theta^{2}}{4}},\quad
{\gamma_{2}}^{2}=\frac{\beta-{\omega_{2}}^{2}\theta}{1-{\omega_{2}}^{2}\frac{\theta^{2}}{4}}.
\end{equation}
In non-commutative phase space, we have $\tilde{\omega_{1}}, \gamma_{1}, \tilde{\omega_{2}}, \gamma_{2}$ which are deformed form of
$\omega_{1}$, and $\omega_{2}$, with $\gamma_{1}=\gamma_{2}=0$. As we know in the harmonic system, $\omega_{1}$, $\omega_{2}$ and also $\gamma_{1}, \gamma_{2}$ describe the stability and non stability of system. In order to have stability, we impose the conditions of $\tilde{\omega_{1}}^{2}>0, \tilde{\omega_{2}}^{2}>0, \gamma_{1}^{2}>0$ and $\gamma_{2}^{2}>0$. So, we have following conditions
\begin{equation}\label{27}
\omega_{1}^{2}>\frac{\beta^{2}}{4},\quad\quad\frac{\omega_{1}^{2}\theta^{2}}{4}<1,\quad \quad \beta>\omega_{1}^{2}\theta,
\end{equation}
and\\
\begin{equation}\label{28}
\omega_{2}^{2}>\frac{\beta^{2}}{4},\quad\quad\frac{\omega_{2}^{2}\theta^{2}}{4}<1,\quad \quad \beta>\omega_{2}^{2}\theta.
\end{equation}
All these conditions guarantee the stability of the model and clarify the parameter of non-commutative space.
\begin{equation}\label{29}
\begin{array}{c}
 \tilde{\Lambda_{1}}=\Lambda_{1}+\frac{\frac{4}{9}\beta^{2}-\frac{9}{64}{\Lambda_{1}}^{2}\theta^{2}}{1+\frac{9}{64}\Lambda_{1}\theta^{2}},\\
 \tilde{\Lambda_{2}}=\Lambda_{2}-\frac{\frac{\beta^{2}}{9}-\frac{9}{16}{\Lambda_{2}}^{2}\theta^{2}}{1-\frac{9}{16}\Lambda_{2}\theta^{2}}.\\
\end{array}
\end{equation}

As we know, the non-commutativity can affect the evaluation of the
universe which plays important role in cosmological constants. The
cosmological constants $\Lambda_{1}$ and $\Lambda_{2}$ in the
corresponding model are responsible for the potential $V(\phi)$
and $V(\sigma)$. The non-commutative phase space influence to such
quantities and change to form of effective cosmological constant
as a $\tilde{\Lambda_{1}}$ and $\tilde{\Lambda_{2}}$. If we
compare the de Sitter cosmology with deformed phase space model in
the late time limit we will arrive to the Eq. (\ref{29})
corresponding to fields of $\phi$ and $\sigma$. As we can see the
 non-commutative parameter $\beta$ plays significant role with respect to $\theta$  in comparing with de Sitter
cosmological constants $\Lambda_{1}$ and $\Lambda_{2}$. Also,
we can see that $\tilde{\Lambda_{1}}$ and $\tilde{\Lambda_{2}}$
will be effective cosmological constants.  In the case of
$\Lambda_{1}=0$ and $\Lambda_{2}=0$ the effective cosmological
constants are,
\begin{equation}\label{30}
\Lambda_{1 eff}=\frac{4}{9}\beta^{2},
\end{equation}
and
\begin{equation}\label{31}
\Lambda_{2 eff}=-\frac{1}{9}\beta^{2},
\end{equation}
where minus sign is coming from the second field.\\
By using the Eq.(\ref{8}) the dynamics of the model is driven,
\begin{eqnarray}\label{32}
\dot{x}_{1}&=&2p_{x_{1}}-{\gamma_{1}}^{2}y_{1}, \quad\quad \dot{y}_{1}=-2p_{y_{1}}-{\gamma_{1}}^{2}x_{1},\nonumber\\
\dot{p}_{x_{1}}&=&{\gamma_{1}}^{2}p_{y_{1}}-2{\tilde{\omega}_{1}}^{2}x_{1}, \quad \dot{p}_{y_{1}}={\gamma_{1}}^{2}p_{x_{1}}+2{\tilde{\omega}_{1}}^{2}y_{1},\nonumber\\
\dot{x}_{2}&=&2p_{x_{2}}-{\gamma_{2}}^{2}y_{2}, \quad\quad \dot{y}_{2}=-2p_{y_{2}}-{\gamma_{2}}^{2}x_{2},\nonumber\\
\dot{p}_{x_{2}}&=&{\gamma_{2}}^{2}p_{y_{2}}-2{\tilde{\omega}_{2}}^{2}x_{2}, \quad \dot{p}_{y_{2}}={\gamma_{2}}^{2}p_{x_{2}}+2{\tilde{\omega}_{2}}^{2}y_{2},
\end{eqnarray}
and the solutions for variables are found,
\begin{equation}\label{33}
\begin{array}{c}
  x_{1}(t)=\eta_{1}e^{-{\gamma_{1}}^{2}t}\sin(2\tilde{\omega_{1}}t+\delta_{1})-\zeta_{1}e^{-{\gamma_{1}}^{2}t}\sin(2\tilde{\omega_{1}}t+\delta_{2}), \\
  y_{1}(t)=\eta_{1}e^{-{\gamma_{1}}^{2}t}\sin(2\tilde{\omega_{1}}t+\delta_{1})+\zeta_{1}e^{-{\gamma_{1}}^{2}t}\sin(2\tilde{\omega_{1}}t+\delta_{2}), \\
\end{array}
\end{equation}

\begin{equation}\label{34}
\begin{array}{c}
  x_{2}(t)=\eta_{2}e^{-{\gamma_{2}}^{2}t}\sin(2\tilde{\omega_{2}}t+\sigma_{1})-\zeta_{2}e^{-{\gamma_{2}}^{2}t}\sin(2\tilde{\omega_{2}}t+\sigma_{2}), \\
  y_{2}(t)=\eta_{2}e^{-{\gamma_{2}}^{2}t}\sin(2\tilde{\omega_{2}}t+\sigma_{1})+\zeta_{2}e^{-{\gamma_{2}}^{2}t}\sin(2\tilde{\omega_{2}}t+\sigma_{2}), \\
\end{array}
\end{equation}
and
\begin{equation}\label{35}
\begin{array}{c}
x_{1}(t)=(a_{1}+b_{1}t)e^{-{\gamma_{1}}^{2}t}+(c_{1}+d_{1}t)e^{-{\gamma_{1}}^{2}t},\\
y_{1}(t)=(a_{1}+b_{1}t)e^{-{\gamma_{1}}^{2}t}-(c_{1}+d_{1}t)e^{-{\gamma_{1}}^{2}t},\\
\end{array}
\end{equation}
where $\beta=2\omega_1$.\\

Also, we rewrite   $x_{2}(t)$ and $y_{2}(t)$ as follows,
\begin{equation}\label{36}
\begin{array}{c}
x_{2}(t)=(a_{2}+b_{2}t)e^{-{\gamma_{2}}^{2}t}+(c_{2}+d_{2}t)e^{-{\gamma_{2}}^{2}t},\\
y_{2}(t)=(a_{2}+b_{2}t)e^{-{\gamma_{2}}^{2}t}-(c_{2}+d_{2}t)e^{-{\gamma_{2}}^{2}t},\\
\end{array}
\end{equation}
where $\beta=2\omega_2$. In commutative case, we  put solutions of
$x_{1}$, $x_{2}$,  $y_{1}$ and $y_{2}$ into the Eqs. (\ref{6}) and (\ref{7}), hence
the corresponding  solutions of scale factor take the following,
\begin{equation}\label{37}
\begin{array}{c}
a^{3}(t)=\frac{9}{4}\eta_{1}\zeta_{1}\sin(2\tilde{\omega_{1}}t+\delta_{1})\sin(2\tilde{\omega_{1}}t+\delta_{2}),\\
a^{3}(t)=-9\eta_{2}\zeta_{2}\sin(2\tilde{\omega_{2}}t+\sigma_{1})\sin(2\tilde{\omega_{2}}t+\sigma_{2}).\\
\end{array}
\end{equation}

Now, we are going to compare $a$, $\phi$ and $\sigma$ in
commutative and non-commutative cases.  First we consider some
conditions on the scale factors for the deformed and non-deformed
phase space. In deformed case we have the following condition,
\begin{equation}\label{38}
\delta=\delta_{1}=\delta_{2},
\end{equation}
and the corresponding deformed scale factor will be as,
\begin{equation}\label{39}
a^{3}(t)=\frac{9}{4}\eta_{1}\zeta_{1}\sin^{2}(2\tilde{\omega}_{1}t+\delta).
\end{equation}

Also for the commutative case (non-deformed) we take the following
condition,
\begin{equation}\label{40}
\chi=\psi_{1}=\varphi_{1},
\end{equation}
and obtain the corresponding scale factor as,
\begin{equation}\label{41}
a^{3}(t)=\frac{9}{16}(\nu_{1}^{2}-\mu_{1}^{2})\sin^{2}(2\omega_{1}t+\chi).
\end{equation}

It would be interesting to note that, the scale factor (\ref{41}) could reach the big bang smoothly from a contracting phase to an expanding one. It means that the big bang singularity might exist, however our solution can be viewed as a singular bounce which is very interesting from cosmological point of view.\\
In the special case, we compare the scale factor from two Eqs.
(\ref{38}) and (\ref{40}) as a non-commutative and commutative respectively.
Here, we first draw the scale factors with respect to time and see
the behavior of such parameters in the Fig. 1. The structure of
scale factors for two systems will be same as sinus form.
But, in non-commutative case, the parameter of $\tilde{\omega}$ as
effective frequency plays important role. This leads us to
consider three cases. In first case we assume that two
non-commutative parameters be equal as $\theta=\beta$, second case
$\beta>\theta$ and third case $\theta>\beta$.  In all cases the parameter
$\beta$ is very sensitive with respect to parameter $\theta$. Also, by
different values of $\theta$ and $\beta$ our results agree with
the Ref. \cite{17}. Likewise, we notice for large values of $t$, our
cosmological model behaves like a de Sitter cosmology.
\begin{figure}
\hspace*{1cm}
\begin{center}
\epsfig{file=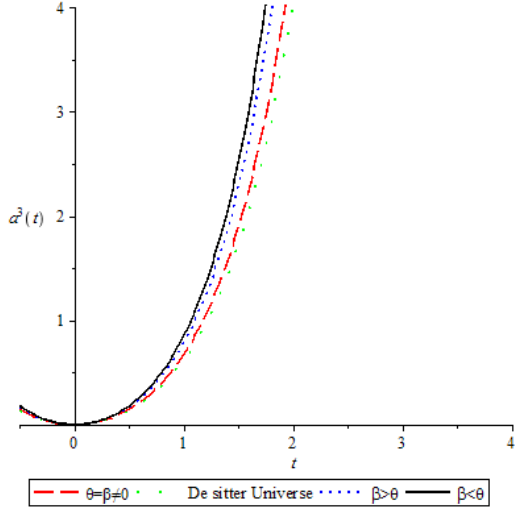,width=7cm}
 \caption{The effect of phase space deformed model with fields $\phi$ and $\sigma$ for the values of cases $\theta=\beta$, $\beta>\theta$ and $\theta>\beta$. We see in this figure for large value of $t$ the behavior is the same.}
\end{center}
\end{figure}
As we know, in the quintom model, we have two fields as $\phi$ and $\sigma$. In that case, the corresponding potentials as $V(\phi)$ and $V(\sigma)$ will be responsible for the vacuum energy. So, we take advantage from the calculation of the scale factor in non-commutative system and obtain fields $\phi$ and $\sigma$ as,
\begin{equation}\label{42}
\begin{array}{c}
\phi(t)=\frac{2\sqrt{2}}{3}\ln\frac{\eta_{1}e^{(-2\gamma_{1}^{2}t)}\sin(2\tilde{\omega_{1}}t+\delta_{1})}{\zeta_{1}\sin(2\tilde{\omega_{1}}t+\delta_{2})},\\
\sigma(t)=\frac{\sqrt{2}}{3}\ln\frac{\eta_{2}e^{(-2\gamma_{2}^{2}t)}\underline{}\sin(2\tilde{\omega_{2}}t+\sigma_{1})}{\zeta_{2}\sin(2\tilde{\omega_{2}}t+\sigma_{2})}.\\
\end{array}
\end{equation}

Now, we consider Eqs. (9) and (\ref{41}) and compare the fields in non-commutative and commutative cases. In that case, we assume same conditions for deformed and non-deformed fields.
In non-commutative case of (\ref{41}) we apply the Eq. (\ref{37}) and following  condition,
\begin{equation}\label{43}
\sigma_{0}=\sigma_{1}=\sigma_{2},
\end{equation}
so, one can obtain the modified fields $\phi$ and $\sigma$ as,
\begin{equation}\label{44}
\phi(t)=\frac{2\sqrt{2}}{3}\ln\frac{\eta_{1}e^{-2\gamma_{1}^{2}t}}{\zeta_{1}},
\end{equation}
and
\begin{equation}\label{45}
\sigma(t)=\frac{\sqrt{2}}{3}\ln\frac{\eta_{2}e^{-2\gamma_{2}^{2}t}}{\zeta_{2}}.
\end{equation}
Also, in commutative case as (19) we apply the Eq. (39) and following  condition,
\begin{equation}
\tau=\psi_{2}=\varphi_{2},
\end{equation}
so, one can obtain the commutative case of fields $\phi$ and $\sigma$ as
\begin{equation}
\phi=\frac{2\sqrt{2}}{3}\ln B, \quad\quad
B=\frac{\nu_{1}+\mu_{1}}{\nu_{1}-\mu_{1}},
\end{equation}
and
\begin{equation}
\sigma=\frac{\sqrt{2}}{3}\ln A, \quad\quad
A=\frac{\nu_{2}+\mu_{2}}{\mu_{2}-\nu_{2}}.
\end{equation}

By comparing Eqs. (46), and (47) with the Eqs. (43), and (44) we say that in the
case of deformed phase space, the fields $\phi$ and $\sigma$ are
function of time. But in non-deformed case the corresponding
fields  are constant and not change with respect to time. So,
non-commutative phase space affect two fields of system such that
the corresponding system  always have a cosmological solution with
any physical condition which is not realized by commutative case.\\
Now, we are going to have some cosmological solutions for inflation
and dark energy  for the corresponding two potentials. In that
case, we take advantage from Noether symmetry and obtain several
conditions for cosmological scenarios. These conditions help us to
investigate the  equation of state and  some condition between
potential and Hubble parameter for the dark energy.
\section{Noether Symmetry Approach}
As mentioned before, for  deformed case we had two parameters as
$\theta$ and $\beta$, they play important role on the fields and
scale factor. In order to see how these parameters affect dark
energy and inflation we have to arrange the equation of state in
usual way. As we know less works attend to the second deformed
parameter $\beta$ for the investigation of dark energy and
inflation, so that always $\beta$ assumed equal one. This condition
lead us to study the dark energy by the original lagrangian
(no deformed form). But, here we take Lagrangian for the general
$\beta$ and $\theta$ and no any condition first apply to these
parameters. Also, we note here the effect of parameter $\beta$  in
cosmology is more than other parameters as $\theta$. So, all above
information give us motivation to employ  Noether symmetry to
investigation of dark energy and inflation for the corresponding
model. So, in that case we have to arrange the deformed lagrangian
as,
\begin{equation}
\tilde{\ell}=-3\dot{a}^{2}a+\frac{1}{2}a^{3}(\dot{\phi}^{2}-\dot{\sigma}^{2})+\frac{\sqrt{2}}{3}a^{3}(\gamma_{1}^{2}\dot{\phi}-\frac{1}{2}\gamma_{2}^{2}
\dot{\sigma})+a^{3}(V(\phi;\theta,\beta)+V(\sigma;\theta,\beta)).
\end{equation}
We assume the following generator of symmetry,
\begin{equation}
X=A\frac{\partial}{\partial a}+B\frac{\partial}{\partial \phi}+C\frac{\partial}{\partial \sigma}+\dot{A}\frac{\partial}{\partial \dot{a}}+\dot{B}\frac{\partial}{\partial \dot{\phi}}+\dot{C}\frac{\partial}{\partial \dot{\sigma}},
\end{equation}
where $A$, $B$ and $C$ are functions of $a$, $\phi$ and $\sigma$.\\
Now, we are going to apply the Noether symmetry approach to the
corresponding potential from fields $\phi$, $\sigma$ and scale
factor $a$. Now, we introduce the the following equations,
\begin{equation}
\dot{A}=\frac{\partial A}{\partial a}\dot{a}+\frac{\partial A}{\partial \phi}\dot{\phi}+\frac{\partial A}{\partial \sigma}\dot{\sigma},
\end{equation}
\begin{equation}
\dot{B}=\frac{\partial B}{\partial a}\dot{a}+\frac{\partial B}{\partial \phi}\dot{\phi}+\frac{\partial B}{\partial \sigma}\dot{\sigma},
\end{equation}
and
\begin{equation}
\dot{C}=\frac{\partial C}{\partial a}\dot{a}+\frac{\partial C}{\partial \phi}\dot{\phi}+\frac{\partial C}{\partial \sigma}\dot{\sigma}.
\end{equation}

According to Noether symmetry and Lie condition as
$L_{X}\tilde{\ell}=0$, we obtain a system of coupled partial
differential equations as,
\begin{equation}
A+2\frac{\partial A}{\partial a}a=0,
\end{equation}
\begin{equation}
\frac{3}{2}A+\frac{\partial B}{\partial \phi}a=0,
\end{equation}
\begin{equation}
\frac{3}{2}A+\frac{\partial C}{\partial \sigma}a=0,
\end{equation}
\begin{equation}
A\gamma_{1}^{2}+\frac{a}{3}\frac{\partial B}{\partial \phi}\gamma_{1}^{2}-\frac{a}{6}\frac{\partial C}{\partial \phi}\gamma_{2}^{2}=0,
\end{equation}
\begin{equation}
-\frac{1}{2}A\gamma_{2}^{2}+\frac{a}{3}\frac{\partial B}{\partial \sigma}\gamma_{1}^{2}-\frac{a}{6}\frac{\partial C}{\partial \sigma}\gamma_{2}^{2}=0,
\end{equation}
\begin{equation}
-6\frac{\partial A}{\partial \phi}+\frac{\partial B}{\partial a}a^{2}=0,
\end{equation}
\begin{equation}
-6\frac{\partial A}{\partial \sigma}-\frac{\partial C}{\partial a}a^{2}=0,
\end{equation}
\begin{equation}
\frac{\partial B}{\partial \sigma}-\frac{\partial C}{\partial \phi}=0,
\end{equation}
\begin{equation}
\frac{\partial B}{\partial a}\gamma_{1}^{2}-\frac{1}{2}\frac{\partial C}{\partial a}\gamma_{2}^{2}=0,
\end{equation}
and
\begin{equation}
3A(V(\phi;\theta)+V(\sigma;\theta))+a(BV_{\phi}(\phi;\theta,\beta)+CV_{\sigma}(\sigma;\theta,\beta))=0.
\end{equation}

The above equation lead us to the following solutions,
\begin{equation}
A=0,
\end{equation}
\begin{equation}
B=c\sigma+c_{1},
\end{equation}
\begin{equation}
C=c\phi+c_{2},
\end{equation}
\begin{equation}
V_{\phi}=0,
\end{equation}
and
\begin{equation}
V_{\sigma}=0.
\end{equation}

By using the conservation law of momentum, one can write following
relation,
\begin{equation}
Q=AP_{a}+BP_{\phi}+CP_{\sigma}, \quad\quad Q=\mu_{0},
\end{equation}
this lead us to following conditions,
\begin{equation}
c=0,\quad\quad c_{1}=\gamma_{1}^{2}, \quad\quad c_{2}=\frac{1}{2}\gamma_{2}^{2},\quad\quad \mu_{0}=0,
\end{equation}
and
\begin{equation}
\gamma_{1}^{2}\dot{\phi}-\frac{1}{2}\gamma_{2}^{2}\dot{\sigma}=\frac{\sqrt{2}}{3}(\frac{1}{4}\gamma_{2}^{4}-\gamma_{1}^{4}).
\end{equation}

By using the equation of motion from action  and corresponding metric background,  one can obtain the following equations,
\begin{equation}
\ddot{\phi}+H(3\dot{\phi}+\sqrt{2}\gamma_{1}^{2})+V_{\phi}(\phi;\theta,\beta)=0,
\end{equation}
\begin{equation}
\ddot{\sigma}+H(3\dot{\sigma}+\frac{\sqrt{2}}{2}\gamma_{2}^{2})+V_{\sigma}(\sigma;\theta,\beta)=0,
\end{equation}
and
\begin{equation}
2\dot{H}+3H^{2}+\frac{1}{2}(\dot{\phi}^{2}-\dot{\sigma}^{2})+\frac{\sqrt{2}}{3}(\gamma_{1}^{2}\dot{\phi}-\frac{1}{2}\gamma_{2}^{2}\dot{\sigma})-(V(\phi;\theta,\beta)+V(\sigma;\theta,\beta))=0.
\end{equation}

Now, we are going to investigate the inflation and also dark energy in this cosmological system. In order to do such process, we use the following relations,
\begin{equation}
H^{2}=\frac{\rho}{3},\quad\quad
\dot{H}=-\frac{1}{2}(\rho+p),\quad\quad \omega=\frac{p}{\rho},
\end{equation}
where $\omega$ is the parameter of equation of state and $H$ is Hubble parameter.
Here, in order to obtain the parameter of equation of state for the investigation of dark energy, we apply the results
of (69) and (70) into Eqs. (71- 74). So, we will arrive to the following equation of state,
\begin{equation}
\omega=\frac{3H^{2}+[\frac{2}{9}(\frac{1}{4}\gamma_{2}^{4}-\gamma_{1}^{4})-2(V(\phi;\theta,\beta)+V(\sigma;\theta,\beta))]}{3H^{2}},
\end{equation}

As we know, in modern cosmology of dark energy, the equation of
state parameter $\omega=\frac{p}{\rho}$ plays an important role,
where $p$ and $\rho$ are its pressure and energy density,
respectively. To accelerate the expansion, the equation of state
of dark energy must satisfy $\omega < \frac{-1}{3}.$ The simplest
candidate of the dark energy is a tiny positive time-independent
cosmological constant $\tilde{\Lambda_1}$ and $\tilde{\Lambda_2}$
for which $\omega=-1$.  With these crossing over condition and
asymptotically behavior of $V (\phi; \theta,\beta)$ and $V (\phi;
\theta, \beta)$ , one concludes that crossing over $-1$ must
happen before reaching the potential minimum asymptotically. This
implies that when crossing over $-1$ occur the deformed fields
$\phi$ and $\sigma$  must continue to run away, this lead us to
have the following conditions for the dark energy,
\begin{equation}
V(\phi)+V(\sigma)<-\frac{3}{2}H^{2},
\end{equation}
and
\begin{equation}
V(\phi;\theta,\beta)+V(\sigma;\theta,\beta)<-\frac{3}{2}H^{2},
\end{equation}
where
\begin{equation}
V(\phi;\theta,\beta)=V(\phi)+\frac{1}{9}\gamma_{1}^{4},
\end{equation}
and
\begin{equation}
V(\sigma;\theta,\beta)=V(\sigma)-\frac{1}{36}\gamma_{2}^{4}.
\end{equation}

We note here, the Noether symmetry helped us to arranged Eqs. (69)
and (70). These equations played important role on the  equation state of dark energy and also cosmic inflation.
\section{Conclusion}

In this paper, we investigated the effects of
non-commutative phase space to quintom cosmological model. Also, we discussed the effects of the deformation
of phase space on $\Lambda_{1}$, $\Lambda_{2}$ and other
cosmological parameters. In the deformation phase space, we obtained some effective  cosmological constants  corresponding to the fields $\phi$ and $\sigma$. So, non-commutative parameters $\theta$ and $\beta$ helped us to investigate the cosmological solution. We obtained plot of the scale factor in terms of time and investigated the effect of non-commutative parameters as $\theta$ and $\beta$ in the corresponding system.
Finally, we took advantage from  non-commutative phase-space and obtained the deformed Lagrangian for the quintom model. In order to discuss the dark energy and inflation, we employed the Noether symmetry. Finally, we compared two non-commutative parameters $\theta$ and $\beta$ for variation of scale factor with respect to time
and shown that the parameter of $\beta$ with respect to $\theta$ is very sensitive.  When this parameter is changed our results will be near to the de Sitter universe which is agreed by Ref. \cite{17}. Also, we achieved some condition  from corresponding potential and investigated the dark energy in commutative and non-commutative case.

\end{document}